\documentclass[10pt,shortpaper,twoside,web] {ieeecolor}

\usepackage{generic}
\usepackage{cite}
\usepackage{amsmath,amssymb,amsfonts}
\usepackage{algorithmic}
\usepackage{graphicx}
\usepackage{textcomp}
\usepackage{psfrag} 
\usepackage{cuted}
\usepackage{float}
\usepackage{graphicx} 
\usepackage{amsmath}  
\usepackage[fancythm,fancybb,morse,ieee]{jphmacros2e} 
\usepackage{array}
\usepackage{bbm}
\usepackage{mathrsfs}
\usepackage{epstopdf}
\DeclareMathAlphabet{\mathpzc}{OT1}{pzc}{m}{it}
\usepackage{graphicx}
\usepackage{amsmath}
\usepackage[fancythm,fancybb,morse,ieee]{jphmacros2e}
\usepackage{caption}
\usepackage{subcaption}
\usepackage{multirow}
\usepackage{csquotes}
\usepackage{url}
\newtheorem{thm}{Theorem}
\newtheorem{lem}{Lemma}

\newtheorem{cor}{Corollary}
\newtheorem{defn}{Definition}

\newtheorem{rem}{Remark}

\def\BibTeX{{\rm B\kern-.05em{\sc i\kern-.025em b}\kern-.08em
		T\kern-.1667em\lower.7ex\hbox{E}\kern-.125emX}}
\begin{document}
	\title{Non-Conservative Data-driven Safe Control Design for Nonlinear Systems with Polyhedral Safe Sets}
	\author{Amir Modares, Bosen Lian, and Hamidreza Modares
		\thanks{ }
		\thanks{A. Modares is with Sharif University, Tehran, Iran (e-mail: amir.modares.81@gmail.com, modares.amir1360@yahoo.com). H. Modares is with Michigan State University, East Lanisng, MI, USA (e-mail: modares@msu.edu). B. Lian is with Auburn University, Auburn, AL, (e-mail:  bzl0098@auburn.edu) }}
	
\maketitle
\thispagestyle{empty} 
\begin{abstract} 
This paper presents a data-driven nonlinear safe control design approach for discrete-time systems under parametric uncertainties and additive disturbances. We first characterize a new control structure from which a data-based representation of closed-loop systems is obtained. This data-based closed-loop system is composed of two parts: 1) a parametrized linear closed-loop part and a parametrized nonlinear remainder closed-loop part. We show that using the standard practice or learning a robust controller to ensure safety while treating the remaining nonlinearities as disturbances brings about significant challenges in terms of computational complexity and conservatism. To overcome these challenges, we develop a novel nonlinear safe control design approach in which the closed-loop nonlinear remainders are learned, rather than canceled, in a control-oriented fashion while preserving the computational efficiency. To this end, a primal-dual optimization framework is leveraged in which the control gains are learned to enforce the second-order optimality on the closed-loop nonlinear remainders. This allows us to account for nonlinearities in the design for the sake of safety rather than treating them as disturbances. This new controller parameterization and design approach reduces the computational complexity and the conservatism of designing a safe nonlinear controller. A simulation example is then provided to show the effectiveness of the proposed data-driven controller. 
\end{abstract}
\begin{IEEEkeywords}
Safe Control, Data-driven Control, Nonlinear Systems, Closed-loop Learning.
\end{IEEEkeywords}

\IEEEpeerreviewmaketitle

\section{Introduction}

\IEEEPARstart{C}ertification of the safety of autonomous control systems is of vital importance for their widespread deployment in real-world applications. Autonomous control systems are, therefore, designed to reach a control task or goal (e.g., optimizing performance) while ensuring their safety. Solving a control problem that accounts for conflicting safety and performance objectives becomes intractable. There are different approaches to overcome this challenge: 1) safety filters \cite{SAfeRL4}-\cite{SAfeRL16} leverage control barrier functions (CBFs) (\cite{SB4}-\cite{SB8}) to correct unsafe actions of a nominal optimal or goal-reaching controller, 2) control merging approaches \cite{merge1} merge a safe feedback controller with a goal-reaching feedback controller (e.g., using R-functions or control sharing \cite{merge1}), and 3) model predictive control (MPC) solves finite-horizon constrained optimal control problems in a receding horizon fashion \cite{MPC1}. 
While CBF-control design for continuous-time (CT) systems can be efficiently designed even for nonlinear systems, CBF-based control design for discrete-time (DT) systems becomes non-convex \cite{safeDTconvex}. The control merging approach learns two separate controllers, i.e., a safe controller and a nominal controller, and merges them to ensure safety while minimizing the intervention with the nominal controller. A challenge is to learn safe controllers for nonlinear systems under parametric uncertainty and disturbances.  Finally, while MPC is a powerful tool for constrained control of linear systems, its computational complexity increase for nonlinear systems. Besides, MPC requires a robust invariant set and its corresponding controller to form a terminal set. Using a linear controller for nonlinear systems can lead to a small-size invariant set. 

Data-based control has recently gained tremendous attention due to its potential to deal with system uncertainties, which are inevitable in real-world control systems. Data-driven safe control design methods can be categorized into direct (i.e., model-free) and indirect (i.e., model-based) learning approaches. The former approach bypasses system identification and leverages the collected data to directly learn a controller by imposing safety constraints. The latter uses the collected data to identify a system model first and then leverages the system model to impose safety constraints. Indirect data-driven safety filters and controllers \cite{Data6}-\cite{Data8}, and data-driven MPC (\cite{MPCd1}-\cite{MPCd5}) have received a surge of attention. However, indirect learning approaches only fit a predictive model to data and do not account for what system model is actually good for the control objectives. That is, they are not control-oriented and can lead to suboptimal performance. 

%As shown in \cite{safeDTconvex}, for DT systems, unless the system is linear and the constraints are also affine, for which it leads to a quadratically constrained quadratic programming (QCQP), the CBF-based approach leads to a non-convex optimization. 
\indent Direct data-driven control has recently been considered for control systems for DT systems 
\cite{Data1}-\cite{Data5}. However, existing results for direct data-driven safe control design of DT systems are mainly limited to linear systems. Recently, data-driven stabilizing controllers have been designed in \cite{Data4,Data3a} for nonlinear DT systems. In one direction, for nonlinear systems in the form of $x(t+1)=AZ(x(t))+Bu(t)$, a nonlinear controller in the form of $u(t)=KZ(x(t))$ is learned to stabilize the closed-loop system. 
This setup assumes that a library of functions is known and capable of describing the dynamics of the system  (i.e., $Z(x)$ is known), which can be obtained by physical laws governing the system dynamics. Letting $Z(x)=[x^T \quad S(x)^T]^T$, the controller is then decomposed into $u(t)=K_1 x(t)+K_2 S(x(t))$, where $K_1$ is learned to locally stabilize the closed-loop system, while $K_2$ is designed to cancel/minimize the nonlinear terms. However, canceling nonlinearities might lead to conservativeness. Nonlinear data-based controllers are learned to design safe controllers in \cite{DataNon1,DataNon2}, based on CBF and contractivity, respectively. Sum-of-squares or semi-definite programming is used to solve the resulting optimization. The nonlinear terms $Z(x)$ are limited to monomials, and nonlinear terms are typically treated as disturbances to be canceled, which can lead to conservatism. 
% These results are limited to local stabilization. 

% In this paper, we show that using this data-based approach based on nonlinearity minimization leads to computational intractability for safe control design of nonlinear systems with polyhedral safe sets. Therefore, it is not straightforward to use this approach to learn a nonlinear safe controller.  

In this paper, we present a computationally tractable data-based direct safe nonlinear controller for nonlinear DT systems with parametric uncertainties and additive disturbances. We first propose a control structure for which its closed-form is composed of a linear part and a Lagrange Remainder (rather than nonlinear terms). We then present a data-based characterization of the closed-loop nonlinear system. We show that extending the data-based nonlinearity cancellation or minimization approach of \cite{Data4} to safe control design leads to computational intractability for safe control design of nonlinear systems with polyhedral safe sets. The computational intractability comes from two sources: 1) a non-convex optimization that must be solved online, and 2) the characterization of the lumped uncertainty caused by the closed-loop representation and the additive disturbances. We leverage primal-dual optimization to propose a novel approach that learns the closed-loop remainder for the sake of safety, rather than treating nonlinearities as disturbances. This is achieved by imposing the optimality conditions of this optimization on the overall closed-loop systems. This approach allows a computationally-tractable characterization of lumped uncertainties, which is considerably less conservative. A simulation example is provided to verify the theoretical results. \vspace{6pt}

\noindent \textbf{Notations and Definitions.} Throughout the paper, $\mathbb{R}^n$ denotes the sets of vectors of real numbers with $n$ elements, $\mathbb{R}^{n \times m}$ denotes  the set of matrices of real numbers with $n$ rows and $m$ columns. Moreover, $Q \ge  0$ ($Q >  0$) denotes that $Q$ is a non-negative (positive) matrix with all its elements being non-negative (positive) real numbers. Moreover, $Q \succeq 0$ indicates that $Q$ is a positive-definite matrix. For two vectors $x_1 \in \mathbb{R}^n$ and $x_2 \in \mathbb{R}^n$, $x_1 \le x_2$ denotes the element-wise inequality. The symbol $I$ denotes the identity matrix of the appropriate dimension. The notation  $\|x\|=\max\{|x_1|,...,|x_n|\}$ is used as the infinity norm of a vector $x=[x_1,...,x_n]^T \in \mathbb{R}^n$, and $x^T$ is the transpose of $x$. Finally, for a matrix $X$, $X_i$ is its $i$-th row, $\|X\|$ is considered as its infinity norm, and $X^{\dagger}$ is its pseudo inverse. 
% \noindent \textbf{Definition 1.} \cite{SetB} A convex and compact set that includes the origin as its interior point is called C-set. \vspace{6pt}

% \begin{defn}
% \cite{SetB}  A polyhedral C-set $\cal{P} (F,g)$ is represented by \vspace{-9pt}
% \begin{align} \label{elip}
% \cal{P} (F,g) = \{ x \in {\mathbb{R}^n}:Fx  \le g\}, 
% \end{align}
% where $g$ is a vector and $F$ is a matrix. 
% \end{defn} \vspace{3pt}
\section{Formalizing the Nonlinear Data-based Safe Control Design}
This section presents the problem formulation and a data-based representation of closed-loop nonlinear systems. \vspace{-7pt}
\subsection{Problem Formulation}
Consider the discrete-time nonlinear system (DT-NS) with dynamics \vspace{-9pt}
\begin{equation}\label{system} 
x(t+1) = {A}Z(x(t)) + Bu(t) + w(t),
\end{equation}
where $x(t) \in \mathbb{R}^n$ is the system's state, $u(t) \in \mathbb{R}^m$ is the control input, and $w \in \cal{W}$ is the additive disturbance. Moreover,
\begin{align}
    Z(x(t)) = \begin{bmatrix} x(t)^\top & S(x(t))^\top \end{bmatrix}^\top \in \mathbb{R}^{N+n},
\end{align}
where $S(x(t)) \in \mathbb{R}^{N}$ is a known dictionary of nonlinear functions that capture the nonlinear part of the system dynamics. Using this representation of $Z(x(t))$, the system parameters can be decomposed into
\begin{align}
    A=[A_1 \quad A_2] \in \mathbb{R}^{n \times {(N+n)}},
\end{align}
where $A_1 \in \mathbb{R}^{n \times {n}}$ and $A_2 \in \mathbb{R}^{n \times {N}}$ are, respectively, the linear and nonlinear parts of the system model. \vspace{3pt}

The following assumptions are made for the DT-NS \eqref{system}. \vspace{3pt}

\begin{assumption}
    The system matrices $A$ and $B$ are unknown while $Z(x)$ is known. Moreover, $Z(0)=0$.
\end{assumption} \vspace{3pt}

\begin{assumption}
   The disturbance set $\cal{W}$ is bounded. That is, $w(t) \in \cal{W}$ where \vspace{-6pt}
   \begin{align}
     \cal{W}= \{w \in \mathbb{R}^n: \|w \| \le h_w \}.
   \end{align}
   \vspace{-15pt}
   % \begin{align}
   %     \cal{W}=\{ w \in {\mathbb{R}^n}: Wx  \le g_w\}
   % \end{align}
   % where $\cal{W} \in \mathbb{R}^{n \times n}$ and $g_w \in \mathbb{R}^n$. where $\cal{W} \in \mathbb{R}^{n \times n}$ and $g_w \in \mathbb{R}^n$. 
\end{assumption} \vspace{3pt}

\begin{assumption}
  The system's safe set is given by a polyhedral C-set $\cal{P} (F,g)$ described by \vspace{-1pt}
  \begin{align} \label{safeset}
      \cal{S}(F,g)=\{ x \in {\mathbb{R}^n}: Fx  \le g\},
  \end{align}
  where $F \in \mathbb{R}^{s \times n}$ and $g \in \mathbb{R}^s$. Here, $s$ is the number of constraints.
\end{assumption} \vspace{3pt}

Since $A$ and $B$ are unknown, a safe controller must be learned using collected data. To collect data, a sequence of control inputs, given as follows, is typically applied to the system \eqref{system}
\begin{align} \label{data-u}
U_0 := \begin{bmatrix} u(0) & u(1) & \cdots & u(T-1) \end{bmatrix} \in \mathbb{R}^{m \times T}.
\end{align}
We then arrange the collected $T+1$ samples of the state vectors as \vspace{-9pt}
\begin{align} \label{data-x1}
X := \begin{bmatrix} x(0) & x(1) & \cdots & x(T) \end{bmatrix} \in \mathbb{R}^{n \times (T+1)}.
\end{align}

These collected samples are then organized as follows 
\begin{align} \label{data-x}
X_0 &:= \begin{bmatrix} x(0) & x(1) & \cdots & x(T-1) \end{bmatrix} \in \mathbb{R}^{n \times T}, \\ \label{data-xx}
X_1 &:= \begin{bmatrix} x(1) & x(2) & \cdots & x(T) \end{bmatrix} \in \mathbb{R}^{n \times T}.
\end{align}

Additionally, the sequence of unknown and unmeasurable disturbances is represented as
\begin{equation}\label{data}
W_0 := \begin{bmatrix} w(0) & w(1) & \ldots & w(T-1) \end{bmatrix} \in \mathbb{R}^{n \times T}.
\end{equation}
which is not available for the control design. \vspace{3pt}

We are now ready to formalize the data-based safe control design problem. The following definition is required. \vspace{6pt}

\begin{defn}
\textbf{Robust Invariant Set (RIS)} \cite{SetB} : The set $\cal{P}$ is a RIS for the system \eqref{system} if $x(0) \in \cal{P}$ implies that $x(t) \in \cal{P} \,\,\, \forall t \ge 0, \,\,\,  \forall w(t) \in \cal{W} $.
\end{defn} \vspace{6pt}

\begin{rem}
   In \cite{Data4}, a nonlinear control policy in the form of
\begin{align} \label{cont}
    u(t)=KZ(x(t))=K_1x(t)+K_2 S(x(t)),
\end{align}
is used to design stabilizing controllers, where $K_1$ is learned so that the linear closed-loop dynamics robustly stabilize the system against nonlinear remainders, and $K_2$ is learned to minimize/cancel the nonlinearities. However, canceling nonlinearities can lead to large control effort and/or conservativeness. In this paper, we propose a different control parameterization and a new nonlinear safe control design approach to avoid conservatism and large control effort. 
\end{rem} \vspace{6pt}

Since $S(x)$ is known (and only their parameters in $A_2$ are unknown), define 
\begin{align} \label{As}
    A_s=\frac{\partial S(x(t))}{\partial t}\big|_{x=0} \in \mathbb{R}^{N \times n},
\end{align}
where the matrix $\frac{\partial S(x(t))}{\partial t}\big|_{x=0}$ is the Jacobian matrix of $S(x)$ being evaluated at the equilibrium point $x=0$. We now propose a controller in the form of
\begin{align} \label{contNew}
    u(t)=K_1x(t)+K_2 Q(x(t)),  
\end{align}
where
\begin{align} \label{Q}
Q(x(t))=S(x(t))-A_s x(t),
\end{align}
is the nonlinearity remainder. Using \eqref{system} and \eqref{contNew}, the closed-loop dynamics  becomes  
\begin{align} \label{clNew}
    x(t+1)=(\bar A_1+BK_1) x(t)+ (A_2+BK_2) Q(x(t))+w(t),
\end{align}
where \vspace{-6pt}
\begin{align} \label{barA}
  \bar A_1=A_1+A_2A_s.
\end{align}
Note that the control law \eqref{contNew}, and, thus, the closed-loop system, is composed of a linear part and a term depending on the Lagrange remainder after linearization. A significant advantage of this controller is that the second term is a closed-loop remainder, for which its dynamics $A_2+BK_2$ will be parameterized, along with the linear part $\bar A_1+BK_1$, in a control-oriented form to ensure that 1) {the linearized dynamics significantly dominate the nonlinear remainders, and 2) the domination occurs for the sake of safety}. Compared to \eqref{cont}, the linear and nonlinear dynamics are now interrelated and will be optimized for the sake of safety. 

Define \vspace{-6pt}
\begin{align} \label{data-z}
    V_0 := \begin{bmatrix} X_0 \\ Q(X_0) \end{bmatrix} \in \mathbb{R}^{(n+N) \times T}.
\end{align}

The following assumptions are required. \vspace{6pt}

\begin{assumption}
   The pair $(\bar A_1,B)$ is stabilizable. Moreover, $Q(x)$ in \eqref{Q} is Lipschitz on the set $\cal{S}(F,g)$, i.e., $\forall x,x_0 \in \cal{S}(F,g)$,  $\|Q(x)-Q(x_0)\| \le L \|x-x_0\| \,\, \text{for some} \,\, L>0$. 
\end{assumption} \vspace{3pt}

\begin{assumption}\label{assumption_5}
The data matrix $V_0$ has full row rank, and the number of samples satisfies $T \geq n+N+1$.
\end{assumption} \vspace{6pt}

\noindent \textbf{Problem 1: Data-based Safe control Design:} 
Consider the system \eqref{system} under Assumptions 1-4. Let the input-state collected data be given by \eqref{data-u} and \eqref{data-x1} and satisfy Assumption 5. Learn a nonlinear safe control policy in the form of \eqref{contNew} to ensure that the safe set \eqref{safeset} is an RIS. \vspace{3pt}

To design a controller that guarantees RIS of the safe set (i.e., to ensure that the system's states never leave the safe set), we leverage the concept of $\lambda$-contractive sets defined next. \vspace{3pt}

\begin{defn} \textbf{Contractive Sets:} \cite{SetB} Given a $\lambda \in (0,1]$, the set  $\cal{P}$ is $\lambda$-contractive for the system \eqref{system} if $x(t) \in \cal{P} $ implies that $x(t+1) \in \lambda \cal{P}, \,\,\, \forall w \in \cal{W}, \,\, \forall t$.  \vspace{3pt}
\end{defn}

Guaranteeing that a set is $\lambda$-contractive not only guarantees that the set is RIS, but also the convergence of the system's states to the origin with a speed of at least $\lambda$. 

\subsection{Data-based Representation of the Closed-loop System}
Inspired by \cite{Data4}, we present a data-based closed-loop representation of the system using the new controller \eqref{contNew}. We also take a slightly different proof than \cite{Data4} to show that the data requirement can actually be more relaxed than assumed in \cite{Data4}. 
\begin{lem}
  Consider the system \eqref{system}. Let the input-state collected data be given by \eqref{data-u} and \eqref{data-x1}, and the state data \eqref{data-x1} are further arranged as \eqref{data-x}-\eqref{data-xx} and \eqref{data-z}. Let the controller be given as \eqref{contNew}. Then,  the data-based closed-loop representation of the system becomes
\begin{align}\label{data-f}
x(t+1) & = (X_1-W_0)G_{K,1} x(t) \nonumber \\
& + (X_1-W_0)G_{K,2}Q(x(t)) + w(t),
\end{align}
where \vspace{-12pt}
\begin{align} \label{datacond}
     & G_K = [G_{K,1} \quad G_{K,2}], \nonumber \\ & 
     V_0 G_{K}=I, \nonumber \\ & 
     K_1=U_0 G_{K,1}, \,\,\,  K_2=U_0 G_{K,2},
\end{align}
with $G_{K,1} \in \mathbb{R}^{T \times n}$
 and $G_{K,2} \in \mathbb{R}^{T \times N}$. Moreover, under Assumption 5, the solution $G_K$ exists and is not unique (and thus can be used as a decision variable).
 % satisfies $V_0 G_{K}=I$, and the control gains are $K_1=U_0 G_{K,1}$ and $K_2=U_0 G_{K,2}$. 
\end{lem} \vspace{3pt}
\noindent \textit{Proof: }
Using the data \eqref{data-x}-\eqref{data-xx}, and the system \eqref{system}, one has
\begin{align}\label{system-data} 
& X_1 = {A}Z_0 + BU_0 + W_0= \nonumber \\ &  \bar A_1 X_0+A_2 Q(X_0)+BU_0+W_0,
\end{align}
where $\bar A_1$ is defined in \eqref{barA}. Using  \eqref{data-z} and multiplying both sides of this equation by
$G_K$, one has
\begin{align}\label{system-cl1} 
& X_1 G_K= [\bar A_1 \quad A_2] V_0 G_K+ BU_0 G_K+ W_0 G_K,
% \nonumber \\ & \quad \quad \,\,\,\,\, =A_l X_0 G_{K,1}+A_n S_0 G_{K,2}+ W_0 G_K.
\end{align}
or equivalently,
\begin{align}\label{system-cl1} 
& (X_1-W_0) G_K=([\bar A_1 \quad A_2] V_0+BU_0) G_{K}.
\end{align}
Using \eqref{datacond} in this equation, the data-based representation of the closed-loop system becomes
\begin{align}\label{system-cl2} 
& (X_1-W_0) G_K=A+BK.
\end{align}
or equivalently,
\begin{align}\label{system-cl2} 
& (X_1-W_0) G_{K,1}=\bar A_1+BK_1, \nonumber \\ & (X_1-W_0) G_{K,2}=A_2+BK_2.
\end{align}
By Assumption 5, a right inverse $G_{K}$ exists such that $V_0 \, G_{K}=I$. {Besides, since the rank of $V_0$ is $n+N$ while at least $n+N+1$ samples are collected, its right inverse $G_k$  exists and is not unique.} Using \eqref{system-cl2} in \eqref{clNew} gives \eqref{data-f}, which completes the proof.  \hfill   $\blacksquare$ \vspace{6pt}

\begin{rem}
 The representation in \cite{Data4} requires that the data matrix \vspace{-6pt}
\begin{align} \label{D0}
   M_0=\begin{bmatrix}
U_0 \\ X_0 \\ Z(X_0)
\end{bmatrix}, 
\end{align}
to be full-row rank, which is mildly stronger than Assumption 5. If $M_0$ is full rank, the system dynamics in the deterministic case can be precisely identified. However, a direct data-dependent approach is a control-oriented approach that does not require an accurate system model. This is related to the data informativeness in \cite{Data1}. 
\end{rem}

\section{Data-based Nonlinear Safe Control using Nonlinearity Minimization}
In this section, we present a data-driven nonlinear safe controller using nonlinearity cancellation. While nonlinearity cancellation is used in \cite{Data4} for local stabilization, we show that it leads to computational intractability in designing safe controllers for systems with polyhedral safe sets.   \vspace{-3pt}
\subsection{Systems with No Disturbance}
We assume in this subsection that $w(t) \equiv 0$. In the next subsection, we show that additive disturbances bring about more computational challenges. The next section will provide a new solution to overcome these challenges.
\begin{thm} \label{firstlem}
    Consider the system \eqref{system} under Assumptions 1, 3-5 with $w(t) \equiv 0$. Let the data-based control gains be given by $K_1=U_0 G_{K,1}$ and  $K_2=U_0 G_{K,2}$. Then, the safe set \eqref{safeset} is $\lambda$-contractive (and thus RIS) using the controller \eqref{contNew} if there exists $P_s$ such that \vspace{-6pt}
\begin{align} \label{cont1}
&{P}_s g  \le \lambda  g-l^d,  \nonumber  \\
&{P}_s F  = F X_1 G_{K,1}, \nonumber \\ 
& V_0 G_{K}=I, \nonumber \\
& P_s \ge 0, 
\end{align}
where $l^d=[l^d_{1},l^d_{2},...,l^d_{s}]^T$ is obtained as
\begin{align} \label{ld}
   & l^d_{i}=\min_{G_{K,2}} \max_{x} F_i X_1 G_{K,2} \, Q(x) \,\,\, \nonumber \\  &{\rm{s}}{\rm{.t}}{\rm{.}} \,\, Fx \le  g.
\end{align} 
Moreover, this controller minimizes the remainder term $X_1 G_{K,2}$ while designing the linear term $X_1 G_{K,1}$ to impose safety. 
\end{thm} \vspace{3pt}

\noindent \textit{Proof:}  To formalize the $\lambda$-contractivity condition, consider the optimization problem 
\begin{align} \label{m1}
&\gamma_i  = \max\limits_{x} {F}_i(\bar A_1+BK_1)x+{F}_i(A_2+BK_2)Q(x) \nonumber \\&
{\rm{s}}{\rm{.t}}{\rm{.}}\,\,\,\,{\rm{ }}Fx \le g, \,\,\, i=1,...,s.
\end{align}
where $(\bar A_1+BK_1)x+(A_2+BK_2)Q(x)$ is the next state of the closed-loop system, obtained from \eqref{clNew} with $w(t)=0$. Therefore, 
$\lambda_i$ represents the maximum of $F_i x(t+1)$ when $F x(t)\le g$. On the other hand, based on Definition 2, $\lambda$-contractivity condition of the closed-loop system requires that if $Fx(t) \le g$, then $F x(t+1) \le \lambda g$. Therefore, the closed-loop system is $\lambda$-contractive if $\gamma_i  \le \lambda g_i, \,\, i=1,...,s$.  Define $f_1(x)=F_i(\bar A_1+BK_1)x$ and $f_2(x)=F_i(A_1+BK_1)Q(x)$. Using the fact that $\max\limits_{x} (f_1(x)+f_2(x)) \le \max\limits_{x} (f_1(x))+\max\limits_{x} (f_2(x))$, if $\max\limits_{x} f_1(x)+\max\limits_{x} f_2(x) \le \lambda g_i$, then $\gamma_i=\max\limits_{x} (f_1(x)+f_2(x)) \le \lambda g_i$ is also satisfied. Therefore, to find a sufficient condition, the linear and the remainder design parts can be separated by optimizing $f_1(x)$ and $f_2(x)$ separately. The nonlinearity cancellation/minimization approach then designs $K_2$ to minimize $f_2(x)$ as 
\begin{align} \label{l}
   & l_i=\min_{K_2} \max_{x} F_i({A_2+B K_2}) \, Q(x) \,\,\, \nonumber \\  &{\rm{s}}{\rm{.t}}{\rm{.}} \,\, Fx \le  g, \,\,\, i=1,...,s.
\end{align} 
Therefore, the $\lambda$-contractivity is guaranteed if $\bar{\gamma_i} \le \lambda g_i, \, i=1,...,s$ with \vspace{-6pt}
\begin{align} \label{model-1}
&\bar{\gamma_i}  = \max\limits_{x} {F}_i(\bar A_1+BK_1)x+l_i \nonumber \\&
{\rm{s}}{\rm{.t}}{\rm{.}}\,\,\,\,{\rm{ }}Fx \le g, \,\,\, i=1,...,s.
\end{align}
where $l_i$ is defined in \eqref{l}.  
The dual of the linear programming optimization \eqref{model-1} is \cite{SetB}
\begin{align}
&\hat \gamma_i  = \min\limits_{\alpha_i} \alpha_i^T \, g + l_i \label{a1} \\&
{\alpha_i}^T F = {F}_i({\bar A_1+B K_1}) \label{a2} \\&
{\alpha_i}^T \ge 0, \,\, i=1,...,s \label{a3}
\end{align}
where $\alpha_i \in \mathbb{R}^s$ is the decision variable of the dual optimization. Define $
{P}_s$ as 
\begin{align} \label{Ps}
{P}_s = \left[ \begin{array}{l}
{\alpha _{1}}^T\\
{\alpha _{2}}^T\\
 \vdots \\
{\alpha _{s}}^T
\end{array} \right] \in {R^{s \times s}}.
\end{align}

$P_s$ is non–negative since $\alpha_i$ is non-negative for all $i=1,...,s$. Moreover, the $\lambda$-contractivity is guaranteed if $\hat \gamma_{i}  \le \lambda g_i $, which leads to ${\alpha _{i}}^T g +l_i \le \lambda g_i \,\,\,\,\, \forall i=1,...,s$. Therefore, using the dual optimization,  $\lambda$-contractivity leads to 
\begin{align} \label{cont2}
&{P}_sg  \le \lambda  g-l,  \nonumber  \\
&{P}_s F  = F (\bar A_1+B \, K_1), \nonumber \\
& P_s \ge 0, 
\end{align}
where $l$ is defined in \eqref{l}. On the other hand, when $w(t) \equiv 0$ in \eqref{system}, the data-based closed-loop representation \eqref{data-f} becomes
\begin{align}\label{data-fnd}
x(t+1) = X_1 G_{K,1} x(t)+ X_1 G_{K,2}Q(x(t)).
\end{align}
Therefore, $\bar A_1+BK_1=X_1 G_{K,1}$ and $A_2+BK_2=X_1 G_{K,2}$. Using these data-based representations in \eqref{cont2} and \eqref{l} completes the proof. \hfill   $\blacksquare$ \vspace{-2pt}
\begin{rem}
Compared to the controller \eqref{cont}, the proposed controller in the form of \eqref{contNew} makes the decision variable $G_{K_1}$ interrelated with the decision variable $G_{K_2}$. To see this, from \eqref{system-cl2} with $W_0=0$, one has $A_2+BK_2=X_1 G_{K_2}$ or $A_2=X_1  G_{K_2}-B K_2$. On the other hand, using $A_2$ in $A_1+A_2A_s+BK_1=X_1  G_{K_1}$, one has \vspace{-3pt}
\begin{align}
    [X_1-BU_0][G_{K_1}-G_{K_2}A_s]=A_1.
\end{align}
Therefore,  $G_{K_1}$ and $G_{K_2}$ are implicitly related, and can be learned to minimize the remainder nonlinearities for the sake of a control objective, rather than canceling nonlinearities. The optimization problem \eqref{l}, however, brings computational challenges. The first challenge is that it is not convex. The second challenge is that it must be solved online as new data becomes available.  \vspace{3pt}
\end{rem} \vspace{-6pt}

% One approach to deal with this challenge is to design the control gain $K=[K_1, \quad K_2]$ and find design $K_1$ for the linear system part $\bar A_1+BK_1$ and $K_2$ to cancel out the nonlinear part $A_2+BK_2$ or, in not possible, treat it as a disturbance that must be minimized. A challenge is that this can be highly-conservative. 

\subsection{Systems with Additive Disturbance}

In the presence of additive disturbance, the closed-loop dynamics \eqref{data-f} can be rearranged as  
\begin{align}\label{eq.DD_closed_loop_final_0}
x(t+1) & =  X_1 G_{K,1} x(t) +  X_1 G_{K,2}Q(x(t)) \nonumber \\
&  -W_0 \big(G_{K,1}x(t)+G_{K,2}Q(x(t))\big)+ w(t)
\end{align}
Since the nonlinear remainder term $X_1 G_{K,2}Q(x(t))$ is considered as a  disturbance to be minimized or canceled, we define the lumped disturbance as a function of both decision variables  $G_{K,1}$ and $G_{K,2}$ as
\begin{align} \label{delta}
   & \delta(t)=X_1 G_{K,2}Q(x(t)) -W_0 \big(G_{K,1}x+G_{K,2}Q(x(t))\big) \nonumber \\ & \quad \quad + w(t).
\end{align}
Apparently, since $w(i), \,\, i=0,...,T-1$ are norm bounded by $h_w$ based on Assumption 2, one has $\|W_0 \| \le T h_w$. Therefore, we define the set 
\begin{align} \label{distSet}
 \cal{W}_W=\{W \in \mathbb{R}^{n \times T}: \|W\| \le Th_w\}. 
\end{align} 
The closed-loop system becomes 
\begin{align}\label{eq.DD_closed_loop_final_1}
x(t+1)  =  X_1 G_{K,1} x(t) +\delta(t).
\end{align}

\begin{cor}
    Consider the system \eqref{system} under Assumptions 1-5 and the control input \eqref{contNew}. Let the data-based control gains be given by $K_1=U_0 G_{K,1}$ and  $K_2=U_0 G_{K,2}$. Then, the safe set \eqref{safeset} is $\lambda$-contractive (and thus RIS) if there exists $P_s$ such that \vspace{-12pt}
\begin{align} 
&{P}_s g  \le \lambda  g-l^{dw}  \nonumber  \\
&{P}_s F  = F X_1 G_{K,1}, \nonumber \\
& V_0 G_{K}=I, \nonumber \\  
& P_s \ge 0,
\end{align}
where $l^{dw}=[l^{dw}_{1},l^{dw}_{2},...,l^{dw}_{s}]^T$ is  obtained as
\begin{align} \label{ld}
   & l^{dw}_{i}=\min_{G_{K,2}} \max_{x} \max_{W_0 \in \cal{W}_W} F_i \delta(t) \,\,\, \nonumber \\  &{\rm{s}}{\rm{.t}}{\rm{.}} \,\, Fx \le  g.
\end{align} 
where $\delta(t)$ is defined in \eqref{delta}. Moreover, this controller minimizes the lumped uncertainty while designing the linear term $X_1 G_{K,1}$ to impose safety.  
\end{cor} \vspace{3pt}
\noindent \textit{Proof:} The proof is similar to the proof of Theorem \ref{firstlem}. \vspace{6pt}
\begin{rem}
    To obtain the optimization problem \eqref{ld}, similar to \cite{Data4}, the nonlinear remainders are treated as disturbances that must be canceled. This optimization is even more challenging than that of \eqref{l}. The uncertainty characterization becomes computationally intractable, and the optimization variables cannot be separated readily.   \vspace{-3pt}
\end{rem}

\section{A Computationally-efficient Data-based Nonlinear Safe Control Design}
% We first assume that $n=N$ (i.e., the number of nonlinear terms is the same as the number of states) and will explain how to relax this condition. 
We now leverage a primal-dual optimization approach to design a computationally efficient safe control solution for a nonlinear system.

% We also assume that the function $Z(x)$ (and thus $Q(x)$) includes a dictionary of $N=nk$ functions for some $k=1,2,...$. When the form of $Z(x)$ is unknown, a dictionary of polynomial functions that satisfy $N=nk$ can be arbitrarily chosen. When the form of $Z(x)$ (and thus $Q(x)$) is known and $N \neq nk$, arbitrary functions can be added to satisfy $N=nk$. 

% Since $N=nk$, let $Q(x)=[Q_1^T(x),...,Q_l^T(x)]^T$ with $Q_j(x) \in \mmathbb{R}^n$. Let also $A_2+BK_2=[A_{21}+BK_{21},...,A_{2l}+BK_{2l}]$ with $A_{2j}+BK_{2j}\in \mmathbb{R}^{n \times n}$. We then design the control gain $K_1$ and $K_{2j}, \, j=1,...,k$ to achieve safety. 

\begin{thm} \label{new}
    Consider the system \eqref{system} with $w(t) \equiv 0$ under Assumptions 1, 3-5 and the controller \eqref{contNew}. Let the data-based control gains be given by $K_1=U_0 G_{K,1}$ and  $K_{2}=U_0 G_{K,2}$.  Let $x_1 \neq 0$ be an arbitrary point inside the safe set and define
    \begin{align} \label{L1L2}
   L_1=\frac{\partial Q(x)}{\partial x}|_{x=x_1} \in \mathbb{R}^{N \times n}, \quad   L_2=\frac{\partial^2 Q(x)}{\partial x^2}|_{x=x_1} \in \mathbb{R}^{Nn \times n}.     
    \end{align}
Then, the safe set \eqref{safeset} is $\lambda$-contractive (and thus RIS) if there exists $P_s \in \mathbb{R}^{s \times s}$ and $P_x \in \mathbb{R}^{s \times n}$ such that 
    \begin{align}
 &   P_s g +P_x \bar x \le \lambda g,  \\
&    P_sF-F X_1G_{K,1}=P_x,  \\
& FX_1G_{K,2}L_1=P_x \\
& -L_2^T A_i^{K_2} \succ 0, \,\, i=1,...,s \\
& V_0 G_{K}=I, \nonumber \\  
& P_s \ge 0,
\end{align}
    where
\begin{align} \label{barx}
    \bar x=\big(x_1+L_1^{\dagger} Q(x_1)\big).
\end{align}
and
\begin{align} \label{AK2}
 A_i^{K_2}= \begin{bmatrix}
      F_i(A_{2}+BK_{2}) & 0& ...&0 \\
       \vdots &  \vdots & ...&  \vdots \\
       0 & 0&...& F_i(A_{2}+BK_{2})
  \end{bmatrix},
\end{align}

\end{thm} \vspace{3pt}
\noindent \textit{Proof:}  The constrained optimization \eqref{m1} can be formulated as an unconstrained optimization problem using the following Lagrange function 
\begin{align} \label{lagn}
  &  L(x,\alpha_i)=\alpha_i^T  (F x -g)-F_i x^+ \nonumber \\& 
  =\alpha_i^T  (F x -g) \nonumber \\& -{F}_i\Big((\bar A_1+BK_1)x+ (A_{2}+BK_{2})Q(x)\Big),
\end{align}
 where $x^+$ is the closed-loop next state given that the current state is $x$, $\alpha_i \in \mathbb{R}^s$, and $\alpha_i \ge 0$ denotes the Lagrange multiplier. Then, for $ \gamma_i$ in \eqref{m1} one has
\begin{align} \label{Lag}
& -\gamma_i=  \min_{x^+} \max_{\alpha_i} L(x,\alpha_i)=\min_{x^+} \max_{\alpha_i} \Big[\alpha_i^T  (F x -g)-F_i x^+ \Big]  \nonumber \\ &
= \min_{x+} \max_{\alpha_i} \Big[\alpha_i^T  (F x -g) 
  -{F}_i\Big((\bar A_1+BK_1)x\nonumber \\&  + (A_{2}+BK_{2})Q(x)\Big) \Big],  \,\, i=1,...,s.
\end{align}
On the other hand, let 
\begin{align} \label{Lag2}
& -\bar \gamma_i= \max_{\alpha_i}  \min_{x+} L(x,\alpha_i) \nonumber \\ &
=  \max_{\alpha_i} \min_{x^+} \Big[\alpha_i^T  (F x -g) 
  -{F}_i\Big((\bar A_1+BK_1)x \nonumber \\& + (A_{2}+BK_{2})Q(x)\Big) \Big], \,\, i=1,...,s.
\end{align}
A sufficient condition for $\lambda$-contractivity of the closed-loop system is to ensure that $\hat \gamma_i \le \lambda g_i, i=1,...,s$ because $\gamma_i \le \bar \gamma_i, \,\, i=1,...,s$. Using the first-order necessary condition and the second-order condition for optimality, the minimum over $x^+$ within the safe set occurs if 
\begin{align} \label{condN}
&    \alpha_i^T  F 
  -{F}_i(\bar A_1+BK_1)-F_i(A_{2}+BK_{2})\frac{\partial Q(x)}{\partial x}=0, \nonumber \\ & 
  -\big[\frac{\partial^2 Q(x)}{\partial x^2}\big]^T A^{K_2}_i \succ 0,
 \end{align}
is satisfied at a point in the safe safe for some $K_1$ and $K_2$, where $A^{K_2}_i$ is defined in \eqref{AK2}. 
Since $K_1$ and $K_2$ can be freely chosen to satisfy these conditions for a given state, we fix $x$ in \eqref{condN} to some $x=x_1$ for which $L_1$ and $L_2$ are defined in \eqref{L1L2}. Then, \eqref{condN} becomes
\begin{align} \label{condN2}
&    \alpha_i^T  F 
  -{F}_i(\bar A_1+BK_1)=F_i(A_{2}+BK_{2})L_1=\beta_i^T, \nonumber \\ & 
  -L_2^T A^{K_2}_i \succ 0, \,\, i=1,...,s.
 \end{align}
where $\beta_i \in \mathbb{R}^{n}$. Using \eqref{condN2} in \eqref{Lag2}, the following condition guarantees the $\lambda$-contractivity
\begin{align} \label{Lag3}
& -\bar \gamma_i \ge  \max_{\alpha_i}  \Big[-\alpha_i^T  g 
  -\beta_i^T \bar x \Big] \ge \lambda g \,\,\, i=1,...,s.
\end{align}
where $\bar x$ is defined in \eqref{barx}. Then, $\lambda$ contractivity is satisfied if 
\begin{align}
 &   \alpha_i^T g +\beta_i^T \bar x \le \lambda g_i, \\
&    \alpha_i^T F-F(\bar A_1+BK_1)=F(A_2+BK_2)L_1=\beta_i^T, \\
& -L_2^T A_i^{K_2} \succ 0, \,\,\, i=1,...,s..
\end{align}
Defining $P_s$ as \eqref{Ps} and 
\begin{align} \label{Ps}
{P}_x = \left[ \begin{array}{l}
{\beta _{1}}^T\\
{\beta _{2}}^T\\
 \vdots \\
{\beta _{s}}^T
\end{array} \right] \in {R^{s \times n}}.
\end{align}
and using $\bar A_1+BK_1=X_1 G_{K,1}$ and $A_{2}+BK_{2}=X_1 G_{K,2}$ completes the proof. \hfill   $\blacksquare$

\begin{cor} \label{newc}
Consider the system \eqref{system} under disturbance. Let Assumptions 1-5 be satisfied and the controller \eqref{contNew} is used with the gains $K_1=U_0 G_{K,1}$ and  $K_{2}=U_0 G_{K,2}$. Then, the safe set \eqref{safeset} is $\lambda$-contractive (and thus RIS) if there exists $P_s \in \mathbb{R}^{s \times s}$ and $P_x \in \mathbb{R}^{n \times n}$ such that 
    \begin{align} \label{main}
 &   P_s g +P_x \bar x+\eta I \le \lambda g,  \\
&    P_sF-F X_1G_{K,1}=P_x,  \\
& FX_1G_{K,2}L_1=P_x \\
& g_m M_x \, T  \big(\|G_{K,1}\| + L  \|G_{K,2}\|+1  \big) \le \eta \\
& -L_2^T A_i^{K_2} \succ 0, \,\, i=1,...,s \\
& V_0 G_{K}=I,  \\  
& P_s \ge 0,
\end{align}  
\end{cor}
where $L_1$ and $L_2$, $\bar x$, and $A^{K_2}_i$ are defined in the statement of Theorem 2, and $g_m=max_i \|F_i\| h_w$ and $M_x$ is the bound on the system's state within the safe set. 

\noindent \textit{Proof:} In the presence of additive disturbance, based on Theorem 2 and using \eqref{eq.DD_closed_loop_final_0}, $\lambda$-contractivity is satisfied if
\begin{align} \label{Lagd}
& -\gamma_i \ge  \max_{\alpha_i} \min_x   L(x,\alpha_i) 
=  \max_{\alpha_i} \min_x  \Big(\alpha_i^T  (Fx-g) \nonumber \\ & 
  -{F}_i (X_1-W_0) G_{K,1} x -F_i  (X_1-W_0) G_{K,2}  Q(x) \nonumber \\ & -F_i w(t) \Big) \ge -\lambda g_i, \quad  \forall w(t) \in \cal{W} \quad   \forall W_0 \in \cal{W}_W, \,\, i=1,...,s
\end{align}

Now, based on the results of Theorem 2, let
\begin{align} \label{condF} 
&\alpha_i^T F  - F_i X_1 G_{K,1}=\beta_i^T, \,\,\, i=1,...,s \nonumber \\
& F_i X_2  G_{K,2}=\beta_i^T, \,\,\, i=1,...,s.
\end{align}
Then, based on the results of Theorem 2, \eqref{Lagd} boils down to 
\begin{align} \label{Lagd2}
&  \min_{\alpha_i}   \Big(\alpha_i^T  g+\beta_i^T \bar x 
  -{F}_i W_0 G_{K,1} x -F_i  W_0 G_{K,2}  Q(x) \nonumber \\ & +F_i w(t) \Big) \le \lambda g_i, \quad  \forall w(t) \in \cal{W} \quad   \forall W_0 \in \cal{W}_W, \,\, i=1,...,s
\end{align}

Using Assumption 4 and since $W_0$ belongs to the set defined in \eqref{distSet}, one has
\begin{align} \label{etab}
  % &   
  % -F_i W_0 G_{K,1}  x-F_i \sum_{j=1}^k W_0 G_{K,2j}  Q_j(x) \le \nonumber \\
  & | -F_i W_0 G_{K,1}  x-F_i  W_0 G_{K,2}  Q(x)| \le \nonumber \\ &  \|F_i\| \|W_0\| \| x\|  \Big(\|G_{K,1}\| + L  \|G_{K,2}\|  \Big) \le \nonumber \\ & T \|F_i\| M_x \, h_w   \Big(\|G_{K,1}\| + L  \|G_{K,2}\|  \Big),
\end{align} 
Therefore, to complete the proof, one needs to guarantee
\begin{align} \label{Lagn2}
  &  \min_{\alpha_i,\beta_i}  \Big(\alpha_i^T g+ \beta_{i}^T  \bar x
  + \nonumber \\ & T \|F_i\| M_x \, h_w   \big(\|G_{K,1}\| + L  \|G_{K,2}\|  \big) \Big) + \|F_i\| h_w  \le \lambda g_i.
\end{align}
Define now $g_m=max_i \|F_i\| h_w$ and the bound on the lumped uncertainty as
\begin{align}
   \delta_c= g_m M_x \, T  \big(\|G_{K,1}\| + L  \|G_{K,2}\|+1  \big).
\end{align}
Then, using \eqref{etab} and the results of Theorem 2, a sufficient condition for the $\lambda$-contractivity to be satisfied is that \eqref{condF} is satisfied and  \vspace{-6pt}
\begin{align}
 &  \max {\eta}_i \nonumber \\&
    {\rm{s}}{\rm{.t}}{\rm{.}}\,\,\,\,{\rm{ }}  \alpha_i^T g+ \beta_{i}^T  \bar x+\eta_i \le \lambda g_i,
    \nonumber \\&T \|F_i\| M_x \, h_w   \big(\|G_{K,1}\| + L  \|G_{K,2}\|  \big)+\|F_i\| h_w \le \eta_i,
\end{align}
is feasible. A compact form of this and using the definitions of $Ps$ and $P_x$ in Theorem 2 and considering the same $\eta$ for all $i=1,...,s$ leads to \eqref{main}. This completes the proof. \vspace{6pt}

\begin{rem}
Theorem 2 and Corollary 2 achieve computational efficiency and resolve conservatism as compared to the control structure used in other nonlinear data-based design methods such as \cite{Data3,Data3a,DataNon2}. This is achieved by  1) proposing a control structure that operates on the nonlinear remainders rather than nonlinear terms. Therefore, it can achieve safety with less conservatism and lower control effort (See also simulation results for comparison), and 2) by accounting for the closed-loop remainder in the optimization and treating it as decision variables (i.e., $\beta_{i}$, which leads to $\bar P_x$) to be decides along with with decisions related to the closed-loop linear part (i.e., $\alpha_i$, which leads to $P_s$) for the sake of safety. This
resolves the challenges of Theorem 1 and Corollary 1. 
\end{rem}

\begin{rem}
    To find the bound $M_x$ on the system state within the safe set, a multi-dimensional interval enclosure of the polyhedral safe set ${F} x(t) \leq {g}$ can be obtained for $x(t)$ as $x(t) \in [\underline{x} \quad \overline{x}]$, where $\underline{x},\overline{x} \in \mathbb{R}^n, \,\, \underline{x} \le \overline{x} $ \cite{Althoff} from which a bound on its norm can be realized depending on $h_s$ and $H_s$ (i.e., $\|x\| \le M_x$ for some $M_x$ depending on $g$ and $F$). 
\end{rem}
% Theorem \ref{new} and Corollary \ref{newc} assume that the number of nonlinear basis functions in $Z(x)$ are the same as the number of states. This assumption can be relaxed as follows. When $n>N$, then we can add $n-N$ zero to the vector $Z(X)$ of basis functions. If $N>n$, we can add zeros to $Z(x)$ to ensure $N=rn$ where $r=2,3,...$ is a natural number. Then, we can rewrite \eqref{beta} as 
% \begin{align} 
% \bar P \bar F= F [A_{21}+BK_{21},....,A_{2r}+BK_{2r}]
% \end{align}
%  where $\bar{P}=[\bar P_1,...,\bar P_r] \in \mathbb{R}^{m \times rm}$ and $\bar{F}=diag(F,...,F) \in \mathbb{R}^{rm \times N}$, $A_2=[A_{21},...A_{2r}$ and $K_2=[K_{21},...K_{2r}]$. This equation can then be broken down to 
%  \begin{align} 
% \bar P_i  F= F (A_{2i}+BK_{2i}), \, i=1,...,r
% \end{align}

\section{Simulation Results}
Consider the following discrete-time system 
% \cite{sim,babak}
\begin{align}
& x_1(t+1) = 0.8 x_1(t) + 0.5 x_2(t)+w(t),  \nonumber \\ &
x_2(t+1) = -0.4 x_1(t) +1.2x_2(t)+x_1^2(t)+x_2^2(t)+u(t).
\end{align}
% where $x_1(t)$ represents the displacement of the carriage from equilibrium, and $x_2(t)$ is the carriage velocity. The input $u(t)$ is the external force. For the system parameters and matrices, readers are referred to \cite{sim}. 
The disturbance $w(t)$ is bounded and satisfies $\|w(t) \| \le 0.05$, and the set parameters are $g=[1,1,1,1]^T$ and 
\begin{align}
    F=\begin{bmatrix}
        0.2 & 0.4 \\
        -0.2 &  -0.4 \\
        -0.15 &  0.2 \\
        0.15 & -0.2
    \end{bmatrix}
\end{align}
The contractility level is chosen as $\lambda=0.95$. 
A controller in the form of $u(t)=K [x_1 \,\, x_2 \,\, x_1^2 \,\, x_2^2]$ is then learned using Theorem 2.
% The control gain is $K=[0.308 \,\, -1.1808 \,\, -0.991  \,\, -1.001]$. Fig. 1 shows the simulation results starting from eight different initial conditions, all of which safely reach to a neighborhood close to the origin. 

% \begin{figure}
% \vspace{-20pt}
% \begin{center}
% \includegraphicswidth  \includegraphics[width=0.52\textwidth, trim=0cm 8.0cm 0cm 4.2cm, clip]{fign.pdf}
% \caption{Simulation result for a nonlinear system. The safe set is shown in red. Eight different trajectories are shown, each starting from the boundary of the safe set. }
% \vspace{-20pt}
% \end{center}
% \end{figure}

% is $\cal{S} = \{x = (x_1, x_2) \in \mathbb{R}^2 \,|\, |x_1| \leq 18, |x_2| \leq 18\}$.  

\section{Conclusion}
We present a data-driven nonlinear safe controller for nonlinear discrete-time systems under disturbances and unknown dynamics. The computational efficiency is preserved using a primal-dual optimization approach and a trick to avoid treating the nonlinearities as pure disturbances. The future work is to extend these results to the case where prior knowledge of the system parameters is available. Incorporating prior knowledge into learning data-based closed-loop characterizations that can be explained by prior knowledge can improve the system's performance.

\bibliographystyle{IEEEtran}

\end{document}